\documentclass[12pt]{article}
\usepackage{epsfig}
\usepackage{array}
\usepackage{url}
\makeatletter
\def\alt{\mathrel{\mathpalette\gl@align<}}
\def\agt{\mathrel{\mathpalette\gl@align>}}
\def\gl@align#1#2{\lower.6ex\vbox{\baselineskip\z@skip\lineskip\z@
\ialign{$\m@th#1\hfil##\hfil$\crcr#2\crcr\sim\crcr}}}
\makeatother

\begin{document}
\begin{flushright}
{\tt hep-ph/0604126}\\
MIFP-06-07 \\
April, 2006 \\
\end{flushright}
\vspace*{2cm}
\begin{center}
{\baselineskip 25pt \large{\bf
Lepton Flavor Violation in Intersecting D-brane Models
} \\

}

\vspace{1cm}

{\large
Bhaskar Dutta and
Yukihiro Mimura
} \vspace{.5cm}

{
\it Department of Physics, Texas A\&M University,
College Station, TX 77843-4242, USA
}
\vspace{.5cm}

\vspace{1.5cm} {\bf Abstract}\end{center}

We investigate lepton flavor violation in the context of intersecting
D-brane models.
We point out that these models have  a  source to generate flavor
violation in the trilinear scalar couplings while the geometry of the
construction leads to degenerate  soft scalar masses for different
generations (as in the minimal supergravity model) at the string
scale. The trilinear scalar couplings are not proportional to the
Yukawa couplings when the $F$-term of the $U$-moduli contribution is
non-zero. Consequently, the lepton flavor violating decay processes
are generated. Only other sources of flavor violations in this model
are the Dirac neutrino Yukawa coupling and the Majorana couplings. The
observed fermion mixings are realized from the ``almost rank 1" Yukawa
matrices, which generate a simple texture for the trilinear scalar
terms. We calculate the branching ratios of $\tau\rightarrow
\mu\gamma$, $\mu\rightarrow e\gamma$ and the electric dipole moment of
the electron in this model. We find that the observation of all the
lepton flavor violating decay processes and the electric dipole moment 
will be able to sort out different flavor violating sources.

\thispagestyle{empty}

\bigskip
\newpage

\addtocounter{page}{-1}

\section{Introduction}
\baselineskip 20pt

The standard model is well established to describe physics below the
weak scale  while it also has a number of parameters, especially in
the flavor sector. Indeed, the patterns of masses and mixings for
quarks and leptons are not very simple and  should be explained in a
fundamental way. Thus, one expects that there exists more fundamental
physics beyond the standard model and the masses and the mixings are
described by some fundamental parameters.

Supersymmetry (SUSY) is the most promising candidate of  new physics.
SUSY models can explain gauge hierarchy problems and suggest gauge
unification such as $SU(5)$ grand unified theory (GUT) with successful
gauge coupling unification in the minimal extension of the SUSY
standard model (MSSM). However, SUSY does not solve  flavor puzzles,
Rather, it increases the number of parameters with flavor indices to
more than  hundred in general. Nevertheless, people are not
discouraged to consider SUSY models since the SUSY breaking parameters
with flavor indices are constrained to suppress flavor changing
neutral currents (FCNC)~\cite{Gabbiani:1988rb}.
Actually, one expects
that the FCNC suppression may be realized by a flavor symmetry, which
may give us a hint of the fundamental physics for flavors.

The minimality of the SUSY breaking parameters is  assumed in  the
minimal supergravity (mSUGRA) mediated SUSY breaking
scenario~\cite{Chamseddine:1982jx}: SUSY breaking scalar masses are
universal and the scalar trilinear couplings ($A$-terms) are
proportional to the Yukawa couplings. The degeneracy of the SUSY
breaking masses corresponds to the $U(3)_L \times U(3)_R$ flavor
symmetry. On the other hand, it is hard to relate the  proportionality
of $A$-terms to such flavor symmetries since the Yukawa couplings
themselves break the symmetries.

The fundamental questions for the flavor sector are the following: 1)
Why do fermions replicate with different masses? 2) Can we explain the
pattern of  the masses and the mixings for quarks and leptons? 3) Why
does the flavor symmetry seem unbroken in the SUSY breaking mass
terms, while the fermion masses break the flavor symmetry? 4) Is the
$A$-term proportionality feasible? When is this proportionality
feasible? How does it look like if it is not proportional?

The intersecting D-brane models
\cite{Berkooz:1996km,Blumenhagen:2000wh,Blumenhagen:2005mu} may answer
such questions. The string theory can describe the particle field
theory as an effective theory, and thus, in principle, it has a
potential to calculate all the parameters by using a few fundamental
parameters. Indeed, the intersecting D-branes are interesting
approaches to construct the standard model. The $N$ stack of D-branes
can form $U(N)$ gauge fields as zero modes of open strings attaching
on the D-branes. Open strings can be attached at the intersection
between the $N$ stack and the $M$ stack of D-branes, and massless
chiral fermions belonging to $(N,\bar M)$ bi-fundamental
representation can appear. Such a situation is very attractive to
obtain quark and lepton fields not only in the standard model
\cite{Cvetic:2001tj} but also in the unified models
\cite{Kokorelis:2002ip,Cvetic:2004ui}.
When the extra dimensions are compactified by torus
such as
$T^6 = T^2 \times T^2 \times T^2$,
the intersecting point of the D-branes can be multiplicated,
and thus the fermions are replicated.
The number of generation is therefore a topological number.

In addition to the realization of the standard-like models, the
effective supergravity Lagrangian is calculable
\cite{Cremades:2003qj,Cvetic:2003ch,Kors:2003wf,Camara:2003ku}. The
Yukawa coupling is obtained as an open string scattering for the
triangle formed by the D-branes. The couplings are described as $e^{-k
A}$, where the triangle area $A$ is formed by  three intersecting
points. For the toroidal compactification models, the Yukawa couplings
are written as theta function of geometrical parameters including
instanton effects. In simple models, the Yukawa matrices are written
in the  factorized form $y_{ij} = x^L_i x^R_j$
\cite{Cremades:2003qj,Chamoun:2003pf}. This originates from a
geometrical reason that the left- and the right-handed fermions are
replicated at the intersecting points on  different tori, and the
Yukawa couplings are given as an exponential form of sum of the
triangle areas. As a result of the factorized form of the Yukawa
coupling, the Yukawa matrices are rank 1, and thus only the 3rd
generation fermions are massive. In order to construct a realistic
model, this issue of Yukawa matrices needs to be resolved and several
possibilities have been considered \cite{Chamoun:2003pf,Dutta:2005bb}.
The Yukawa matrices can be hierarchical when the Yukawa matrices are
``almost rank 1" by including higher order effects or quantum
corrections. The exponential suppression of the Yukawa coupling is
also available to obtain hierarchical masses.
In Ref.\cite{Dutta:2005bb}, we have discussed that the observed patterns
of fermion mixings can be easily reproduced if the Yukawa matrices are
almost rank~1.

The K\"ahler metrics of the zero modes can also be calculated as
string scattering amplitudes \cite{Cvetic:2003ch} in terms of the
moduli fields: dilaton $S$, K\"ahler moduli $T$, and complex structure
moduli $U$. The K\"ahler metrics are diagonal for the zero modes, and
the metrics for the bifundamental fields are determined by the brane
configuration parameters such as the relative angles of the D-branes.
Since the relative angles are common  when the fermions are replicated
at the intersection of the D-branes, the K\"ahler metrics are flavor
invisible. Consequently, the SUSY breaking scalar masses are same for
different generations. So, the flavor symmetry of the scalar masses
can originate from the brane geometry.

When the K\"ahler metrics remain same for each generation, the
K\"ahler connection parts (which are the derivatives of K\"ahler
metric) of $A$-terms are common. Thus, the non-proportional part of
the $A$-term is only the derivative of the Yukawa coupling. The Yukawa
couplings which are given as theta functions depend only on the
$U$-moduli \cite{Camara:2003ku}, neither on the $S$ nor the
$T$-moduli. As a result, the trilinear scalar couplings are
proportional to the Yukawa coupling when the $F$ component of the
$U$-moduli is zero. If $F^U\neq 0$, the non-proportional part of
$A$-term is acquired, which is proportional to the derivative of the
Yukawa matrices.

In this paper, we emphasize the  degeneracy of the SUSY breaking mass
terms and the $U$-moduli contribution of the trilinear scalar
couplings. These contributions are related to the flavor violation and
we will study the lepton flavor violation (LFV) processes since the
flavor violation in the lepton sector produce much more stringent
constraint rather than in the quark sector. The LFV processes, such as
$\mu \rightarrow e \gamma$, $\tau \rightarrow \mu \gamma$ and $\tau
\rightarrow e \gamma$, are not yet observed, but we have bounds on the
branching ratios of these decay modes \cite{Brooks:1999pu}.
The  observation of these decay modes would provide   information of
 flavor violation in new physics. It is pointed out that the
processes are accessible for  SUSY models \cite{Borzumati:1986qx}. In
the mainstream of  theoretical calculations, the flavor violation in
the SUSY breaking parameters is assumed to be absent at the cutoff
scale in the mSUGRA. The source of flavor violation originates from
the Yukawa couplings such as the Dirac neutrino  and the Majorana
couplings. The SUSY breaking parameters at the weak scale can acquire
flavor violation through renormalization group equations (RGE).
Indeed, the off-diagonal elements of the left- and right-handed
slepton mass matrices are generated, and the slepton-gaugino loop
diagram provides the LFV processes. In the intersecting D-brane
models, the flavor degeneracy of the slepton masses at the cutoff
scale is realized naturally and the $U$-moduli contribution of
$A$-terms can be one of the major sources of flavor violation. We will
calculate the branching ratios of different LFV decays and the
electric dipole moment (EDM) of the electron, and study whether we can
learn the origins of  flavor violation from the forthcoming
experiments.

This paper is organized as follows: In section 2, we will study the
low energy effective action of the zero modes for matter fields in
intersecting D-brane models. In section 3, the realization of neutrino
mixing angles are obtained in the context of ``almost rank 1 Yukawa
matrix". In section 4, we investigate the sources of LFV. In section
5, we will calculate branching ratios of the LFV decays and the EDM of
the electron, and compare the results with different setups of flavor
violation. Section 6 is devoted to conclusions and discussions.

\section{Effective action in the intersecting D-brane models}

Our purpose is to deal with the flavor physics and study its
phenomenological implications in the intersecting D-brane models. We
will consider models with intersecting D6-branes in the type IIA
theory \cite{Blumenhagen:2005mu}, which may be equivalent to the
models with magnetized D9-branes and D5-branes in the type IIB theory.
One can also apply our work to the D7-branes in the type IIB theory in
which supersymmetry breaking soft terms can arise from 3-form fluxes
\cite{Camara:2003ku}. In this section, we will explore the flavor
sector of the models without concentrating on the details of any
individual model.

The MSSM-like models can be constructed easily by introducing three
sets of intersecting D-branes. For example, in the type IIA
orientifold models with $T^6/Z_2\times Z_2$ and with the intersecting
D6-branes, the $N$ stack of D-branes can form $SU(N/2)$ gauge fields.
Massless chiral fermions belong to the $(N/2,\bar M/2)$ bi-fundamental
representation can appear at the intersection between the $N$ stack
and the $M$ stack of D-branes. So, introducing $a,b,c$ branes for
$U(4)_c$, $U(2)_L$ and $U(2)_R$ respectively, we can obtain
Pati-Salam-like model \cite{Pati:1974yy} with quark and lepton fields
\cite{Cvetic:2004ui}. The $SU(4)_c$ and $SU(2)_R$ symmetries are
broken to $SU(4)_c \rightarrow SU(3)_c \times U(1)_{B-L}$ and $SU(2)_R
\rightarrow U(1)_R$ by brane splitting \cite{Cvetic:2004nk}, and
$U(1)_{B-L} \times U(1)_R$ is broken down to $U(1)_Y$ by the Higgs
mechanism. When the branes are parallel to the orientifolds, the $USp$
gauge symmetry arises. The $SU(2)$ gauge symmetry in the standard
model can  originate from the $USp$ brane. In order to eliminate the
RR tadpole, extra branes are often needed and they may form hidden
sectors \cite{Cvetic:2003yd}.

Since the extra dimensions are compactified to
$T^6 = T^2 \times T^2 \times T^2$,
the D-branes intersects multiple times
and the generations of the fermions are replicated.
The intersection numbers $I_{ab}$ for $D_a$ and $D_b$ branes
are topological invariant
and can be given by the wrapping numbers and the total magnetic fluxes
$(n_a^r,m_a^r)$,
\begin{equation}
I_{ab} = \prod_{r=1}^3 (n_a^r m_b^r - m_a^r n_b^r),
\end{equation}
where $r$ represents  index of each torus. In a simple choice to
obtain three generations, the left-handed matter $\Psi^{ab}$ and the
right-handed matter $\Psi^{ca}$ are often replicated on  different
tori.

The important implication of the family replication
is that the K\"ahler metric is flavor diagonal:
\begin{equation}
K = \hat K(M,\bar M) +  K_{ab}(M,\bar M) \Psi_i^{ab} \bar \Psi_i^{ab} +
K_{ca}(M,\bar M) \Psi_i^{ca} \bar \Psi_i^{ca} + \dots ,
\end{equation}
where  $M$ stands for the $S$, $T$ and $U$ moduli and the index $i$ is
for the flavor index. In addition to the flavor diagonal nature, the
K\"ahler metric for $\Psi^{ab}$ does not depend on  flavor indices for
$I_{ab} = I_{ca} = 3$. The K\"ahler metric $K_{ab}(M,\bar M)$ is
determined by the relative angles $\theta_{ab}^r$ of the D-branes
\cite{Cvetic:2003ch,Kors:2003wf,Camara:2003ku}:
\begin{equation}
K_{ab} \propto e^{\phi_4}
\prod_r (U_r+\bar U_r)^{-\nu_r} \sqrt{\frac{\Gamma(1-\nu_r)}{\Gamma(\nu_r)}} \,.
\end{equation}
where $\phi_4$ is a 4-dimensional dilaton and $\nu_r =
\theta_{ab}/\pi$, which are functions of $S$ and $T$ moduli. The
moduli dependence of the K\"ahler metric is determined by the
geometrical parameters, and thus the metric is flavor invisible.
Therefore, the SUSY breaking scalar mass  for the left-handed matter
is given below \cite{Brignole:1997dp} and it has the flavor
degeneracy:
\begin{equation}
m^2_{ab} = m_{3/2}^2 + V_0 -
       \sum_{M,N} \bar F^{\bar M} F^N \partial_{\bar M} \partial_N \log K_{ab} \,,
\end{equation}
where $m_{3/2}$ is a gravitino mass and $V_0$ is the vacuum
expectation value of the scalar potential. The scalar mass squared
$m_{ca}^2$ for the right-handed matter can be similarly written and
can also have the flavor degeneracy. Since the relative angles for
$ab$ and $ca$ can be different, the scalar masses are not necessarily
universal for  different representations of matter. We note that the
flavor degeneracy can be broken when 2+1 decomposition of the
generation is considered, for example, $I_{ab}= 2$ and
$I_{ab^\prime}=1$, where $b^\prime$ is a orientifold reflection of the
brane $b$. We will choose $I_{ab} = I_{ca} = 3$ from now on  and there
is a  $U(3)_L \times U(3)_R$ flavor symmetry in the SUSY breaking
scalar mass terms at the string scale. We emphasize that such a
mSUGRA-type flavor structure can be obtained by a geometrical setup of
the D-branes.

\begin{figure}[t]
 \center
 \includegraphics[width=10cm]{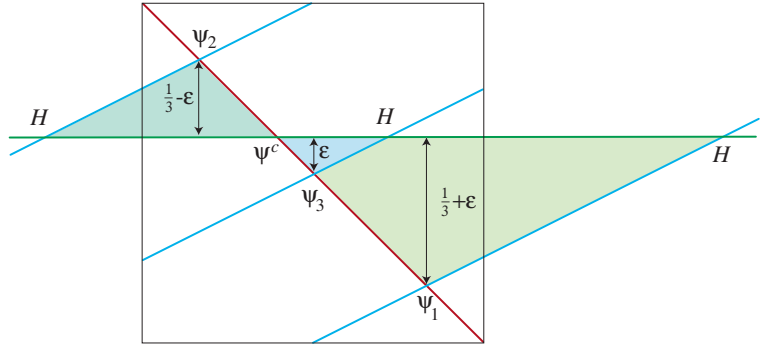}
 \caption{A sketch of brane intersections on a torus. }
\label{schetch02}
\end{figure}

The Yukawa coupling $\Psi^{ab} \Psi^{bc} \Psi^{ca}$ is induced by the
three-point open string scattering. When the left-handed matter
$\Psi^{ab}$ and the right-handed matter $\Psi^{ca}$ are replicated in
different tori, the Yukawa coupling is factorized:
\begin{equation}
Y_{ij} = x_i^L (U_{1}) x_j^R (U_{2}).
\end{equation}
The $x_i^{L,R}$ can be written by theta function
\cite{Cremades:2003qj} with some geometrical parameters such as
$\varepsilon$ shown in the Fig.1. Naively, these are given by $e^{-k
A}$ where $A$ is the area of the triangle formed by the branes. The
Yukawa couplings do not depend on the $S$ and $T$ moduli but depend on
the complex structure moduli $U$.

When the Yukawa coupling is factorized, the matrix is rank 1 and
consequently $U(2)_L \times U(2)_R$ flavor symmetry remains and the
fermions of 1st and 2nd generation are massless. Surely such a
situation is not viable, and there exists  several discussion on this
issue \cite{Chamoun:2003pf}. For example, we have suggested that the
multi-point function of the string scattering including extra branes
can increase the rank of the Yukawa matrix \cite{Dutta:2005bb}. In
this paper, we do not specify how to increase the rank, but we assume
the factorizability of the Yukawa coupling at the leading order since
this assumption leads to  interesting phenomenological implications
which we will see in the next section.

The scalar trilinear coupling ($A$-term) is given as \cite{Brignole:1997dp}
\begin{equation}
A_{ij} = F^M \left[\left(\hat K_M
         - \partial_M \log(K_{ab} K_{bc} K_{ca})\right) Y_{ij}
+ \partial_M  Y_{ij} \right] ,
\end{equation}
and the coupling is proportional to the Yukawa coupling if $F^U = 0$.
However, when $F^U \neq 0$, the flavor violation is generated,
\begin{equation}
F^{U_r} \partial_{U_r} Y_{ij} = F^{U_{1}} \dot x_i^L x_j^R + F^{U_{2}} x_i^L \dot x_j^R,
\label{U-moduli}
\end{equation}
where $\dot x$ stands for the derivative by the $U$ moduli. The $U$
moduli contribution in the $A$-term can be the source of  flavor
violation. In this paper, we will emphasize the effect of the $U$
moduli contribution.

\section{Application of ``almost rank 1 Yukawa matrices"}

We will first discuss the mixing angles for neutrino oscillation in
the context of ``almost rank 1 Yukawa matrix" \cite{Dutta:2005bb}.

The Yukawa matrix for the charged-leptons is given as $Y_e = Y_0 +
\delta Y$ where $Y_0$ is the rank 1 matrix, and $\delta Y$ is a small
correction to generate electron and muon masses.
The rank 1 matrix can be expressed as
\begin{equation}
Y_0 = \left(
       \begin{array}{c} c_1 \\ b_1 \\ a_1
       \end{array}
      \right)
      \left(
       \begin{array}{ccc} c_2 & b_2 & a_2
       \end{array}
      \right)
= \left(
   \begin{array}{ccc}
     c_1 c_2 & c_1 b_2 & c_1 a_2 \\
     b_1 c_2 & b_1 b_2 & b_1 a_2 \\
     a_1 c_2 & a_1 b_2 & a_1 a_2
   \end{array}
  \right),
\label{Yukawa0}
\end{equation}
and $a_i,b_i,c_i$ can be given by theta function
\cite{Cremades:2003qj}. Note that $a_i,b_i,c_i$ can be rotated to be
real by field redefinitions.

Now let us work in  a basis where the mass matrix of the light
neutrino is diagonalized. Then the diagonalizing matrix of $Y_e$ is
the MNSP (Maki-Nakagawa-Sakata-Pontecorvo)
matrix: $U_{\rm MNSP} = U_L^{e*}$, where $U_L^e Y_e
U_R^{e\dagger} = Y_e^{\rm diag}$.
The $3\times 3$ unitary diagonalization matrix has three mixing
angles, and those angles may be generically large since $a_i,b_i,c_i$
are all order one parameters. However, one of the three mixing angles
of $U_{L(R)}$ is unphysical in the limit $\delta Y \rightarrow 0$
since 1st and 2nd generation masses are equal to be zero and the
$U(2)_L\times U(2)_R$ flavor symmetry remains unbroken. The small
correction, $\delta Y$, eliminates the degeneracy and the mixing angle
of $U_{L(R)}$ is fixed. Namely, the  two mixing angles in $U_{L(R)}$
are generically large and  one mixing angle is determined by a small
correction $\delta Y$. For example,
when the small correction is
$\delta Y = {\rm diag}(0,0,\epsilon)$, the Yukawa coupling becomes
rank 2 and the eigenvector for the zero eigenvalues is $\propto
(b_1,-c_1,0)$ for $U_L$. As a result, one can find that $U_{e3}(= \sin
\theta_{13})$ is exactly zero in this example. The small correction
needs to be more realistic to generate the  electron mass, and then
$U_{e3}$ can acquire a small  non-zero value. Consequently, the two
large mixings for solar and atmospheric neutrinos and the  small
mixing for $\theta_{13}$ is elegantly realized in this scheme.


The approximate diagonalization matrix $U_L^0$ is given as
\begin{equation}
U_L^0 = \left( \begin{array}{ccc}
                 \cos \theta^L_s & - \sin\theta^L_s & 0 \\
                 \cos \theta^L_a \sin\theta^L_s & \cos\theta^L_a \cos\theta^L_s & -\sin\theta^L_a \\
                 \sin \theta^L_a \sin\theta^L_s & \sin\theta^L_a \cos\theta^L_s & \cos\theta^L_a
               \end{array}
        \right),
\end{equation}
where $\tan\theta_s^L = c_1/b_1$ and $\tan\theta_a^L =
\sqrt{b_1^2+c_1^2}/a_1$. The right-handed  $U_R^0$ can be also
described similarly. Then the MNSP matrix can be written as
\begin{equation}
U_{\rm MNSP} = V^{e*}_L U_L^0\,,
\label{MNSP-1}
\end{equation}
where $V_L^e$ is a diagonalizing matrix of
$U_L^0 Y_e U_R^{0\rm T}= Y_0^{\rm daig} + U_L^0 \delta Y U_R^{0\rm T}$.

In the quark sector, the CKM (Cabibbo-Kobayashi-Maskawa)
matrix is written as $V_{\rm CKM} = U^u_L
U^{d\dagger}_L$, where the unitary matrices are $U^{u,d}_L Y_{u,d}
U^{u,d\dagger}_R = Y_{u,d}^{\rm diag}$. The Yukawa matrices are given
as $Y^{u,d}_{ij} = x_i^L x_j^{R(u,d)} + \delta Y^{u,d}_{ij}$.
In a similar way in the  charged lepton sector, the unitary matrices can
be written in the form $U^{u,d}_L = V_L^{u,d} U_L^{0}$. Since the
left-hand part $x_i^L$ is common for both up- and down-type quarks,
the large mixings in $U^{u,d}_L$ get cancelled, and the CKM mixings
are small: $V_{\rm CKM} = V^u_L V^{d\dagger}_L$.
Since the up-type quarks are more hierarchical than the down-type
quarks, the CKM matrix is expected to be $V_{\rm CKM} \simeq
V_L^{d\dagger}$. If we have quark-lepton unification, we have a
relation $V_L^d \simeq V_L^e$. Then the MNSP matrix is
\begin{equation}
U_{\rm MNSP} \simeq V_{\rm CKM}^{\rm T} U_L^0.
\end{equation}
This type of MNSP matrix is surveyed as an ansatz in Ref.\cite{Giunti:2002ye}.

Once the MNSP matrix is given in the form Eq.(\ref{MNSP-1}),
we obtain the mixing angles for neutrino oscillation as follows \cite{Dutta:2005bb}:
\begin{equation}
\sin \theta_{13} \simeq \sin \theta_a^L \sin\theta_{12}^e,
\quad
\theta_{\rm atm} \simeq \theta_a^L,
\quad
\theta_{\rm sol} \simeq \theta_s^L \pm \theta_{13} \cot \theta_{\rm atm} \cos \delta_{\rm MNSP}\,,
\label{neutrino-mixing}
\end{equation}
where $\theta_{12}^e$ is a mixing angle in $V_L^e$,
and $\theta_{13}^e$ and $\theta_{23}^e$ are neglected since they are expected to be small
as in the quark sector.
The atmospheric neutrino mixing is almost maximal and the solar mixing
angle is large but not maximal since $c_1 \alt b_1 \alt a_1$ and
$\tan\theta_a^L = \sqrt{b_1^2+c_1^2}/a_1$, $\tan\theta_s^L = c_1/b_1$.

The smallness of the neutrino masses are explained by the seesaw
mechanism \cite{Minkowski:1977sc}. We note that the large mixing
angles between the charged-lepton and the neutrino Dirac Yukawa
coupling can also get cancelled as in the quark sector. Thus, the
favorable situation is when the $SU(2)_L$ triplet Majorana part is
dominant in the type II seesaw \cite{Schechter:1980gr}
\begin{equation}
m_\nu^{\rm light} = M_L - M_\nu^D M_R^{-1} M_\nu^{D\rm T}.
\end{equation}
The Majorana couplings for both left- and right-handed leptons
\begin{equation}
\frac12 f_L \ell \ell \Delta_L +
\frac12 (f_R^{ee} e^c e^c \Delta_R^{--} + f_R^{\nu\nu}\nu^c \nu^c \Delta_R^0
          + \sqrt2 f_R^{\nu e} \nu^c e^c \Delta_R^-),
\label{Majorana}
\end{equation}
can be generated by multi-point functions in each torus \cite{Dutta:2005bb}.

\section{Possible sources of lepton flavor violation}

In this section, we will describe the sources of the LFV processes
such as $\mu \rightarrow e \gamma$ and $\tau \rightarrow \mu\gamma$.

\begin{figure}[t]
 \center
 \includegraphics[width=6cm]{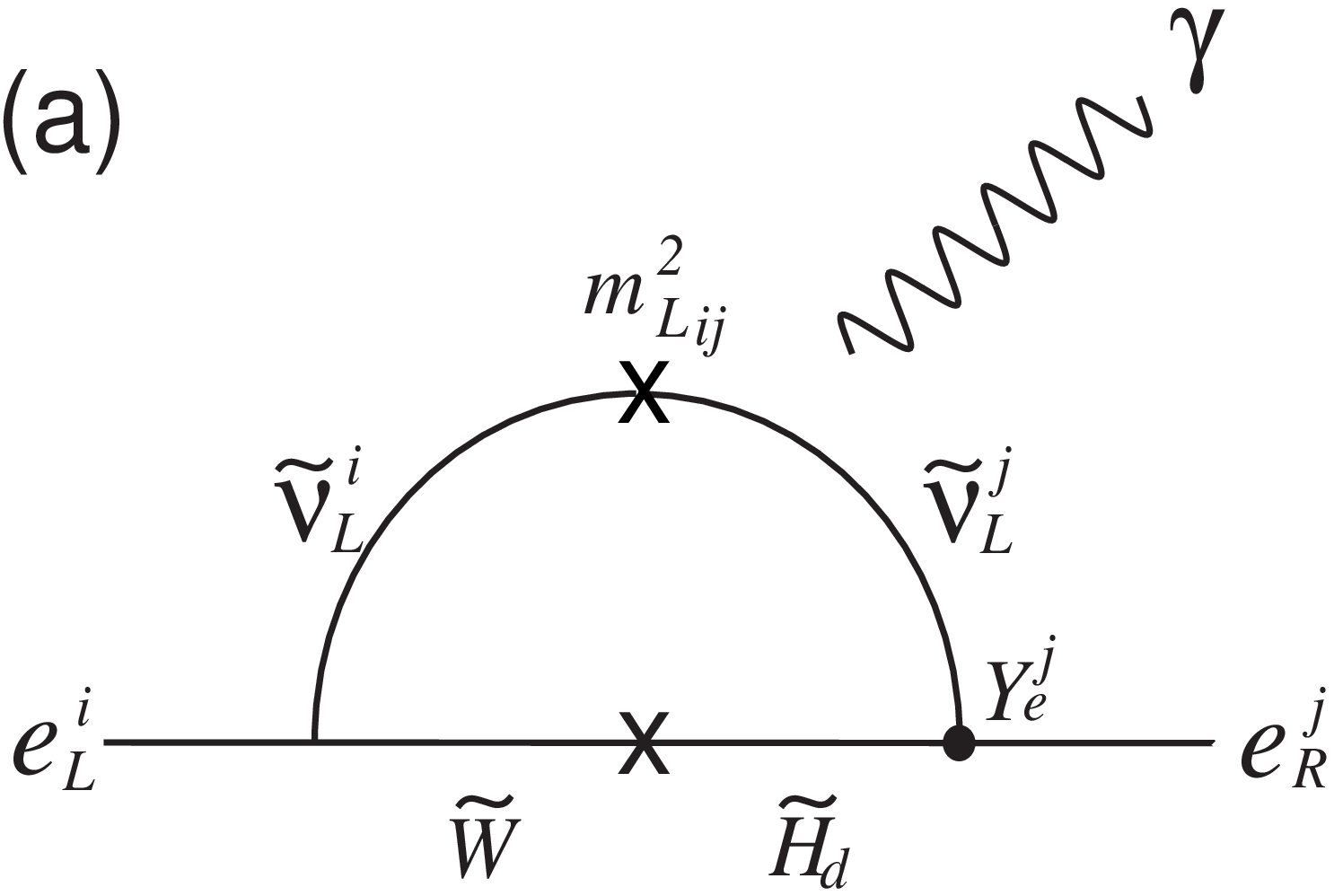}
 \includegraphics[width=6cm]{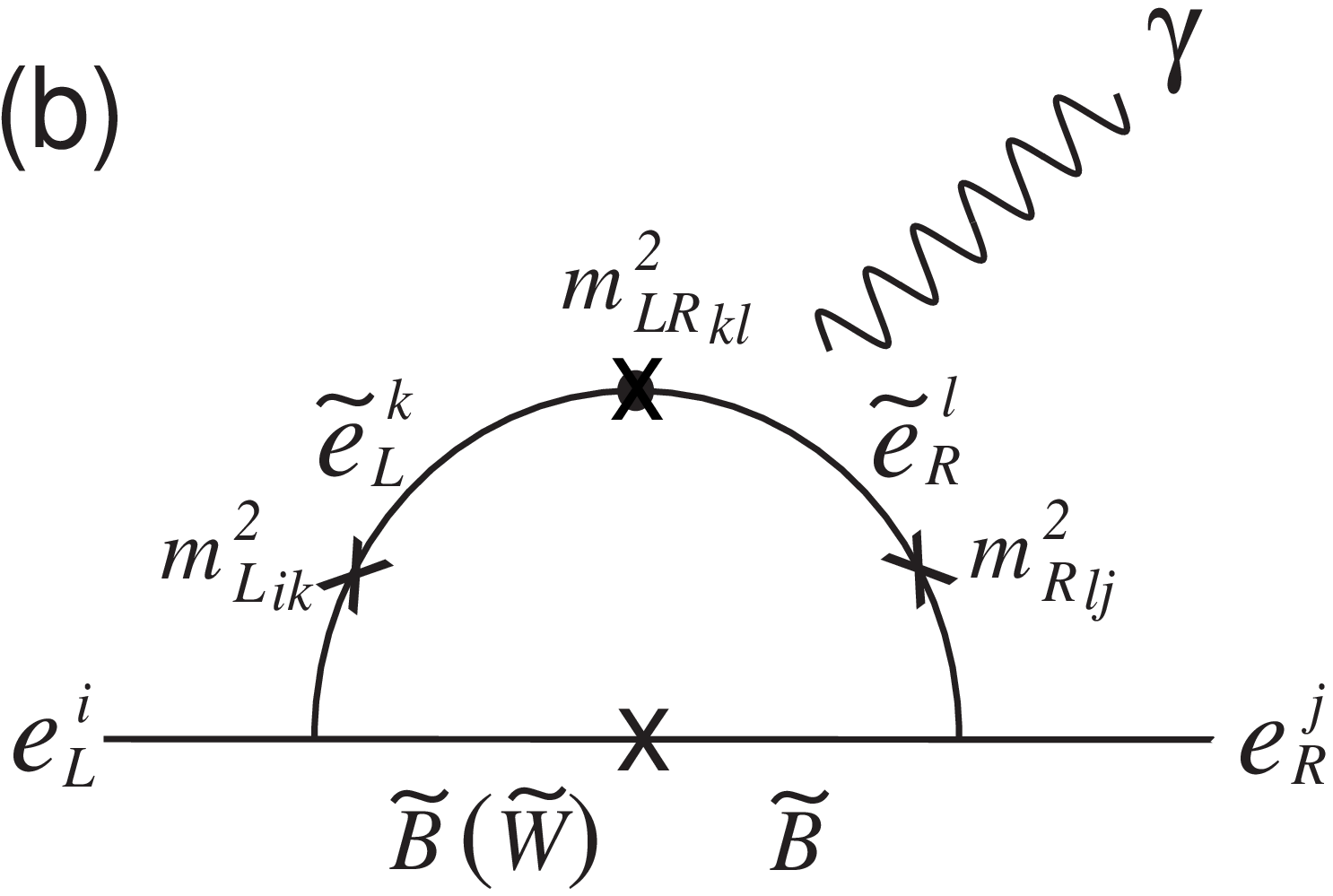}
 \includegraphics[width=6cm]{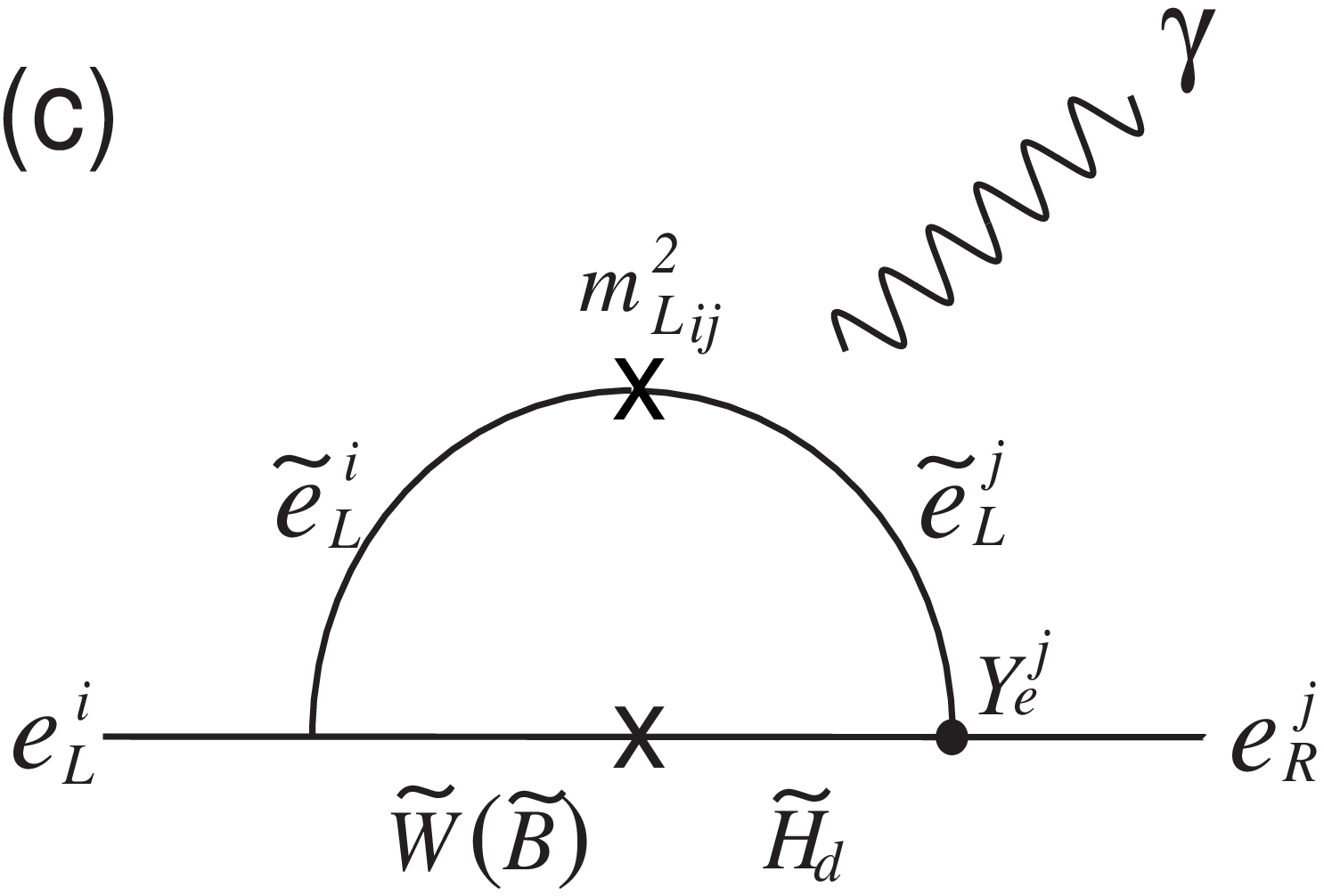}
 \includegraphics[width=6cm]{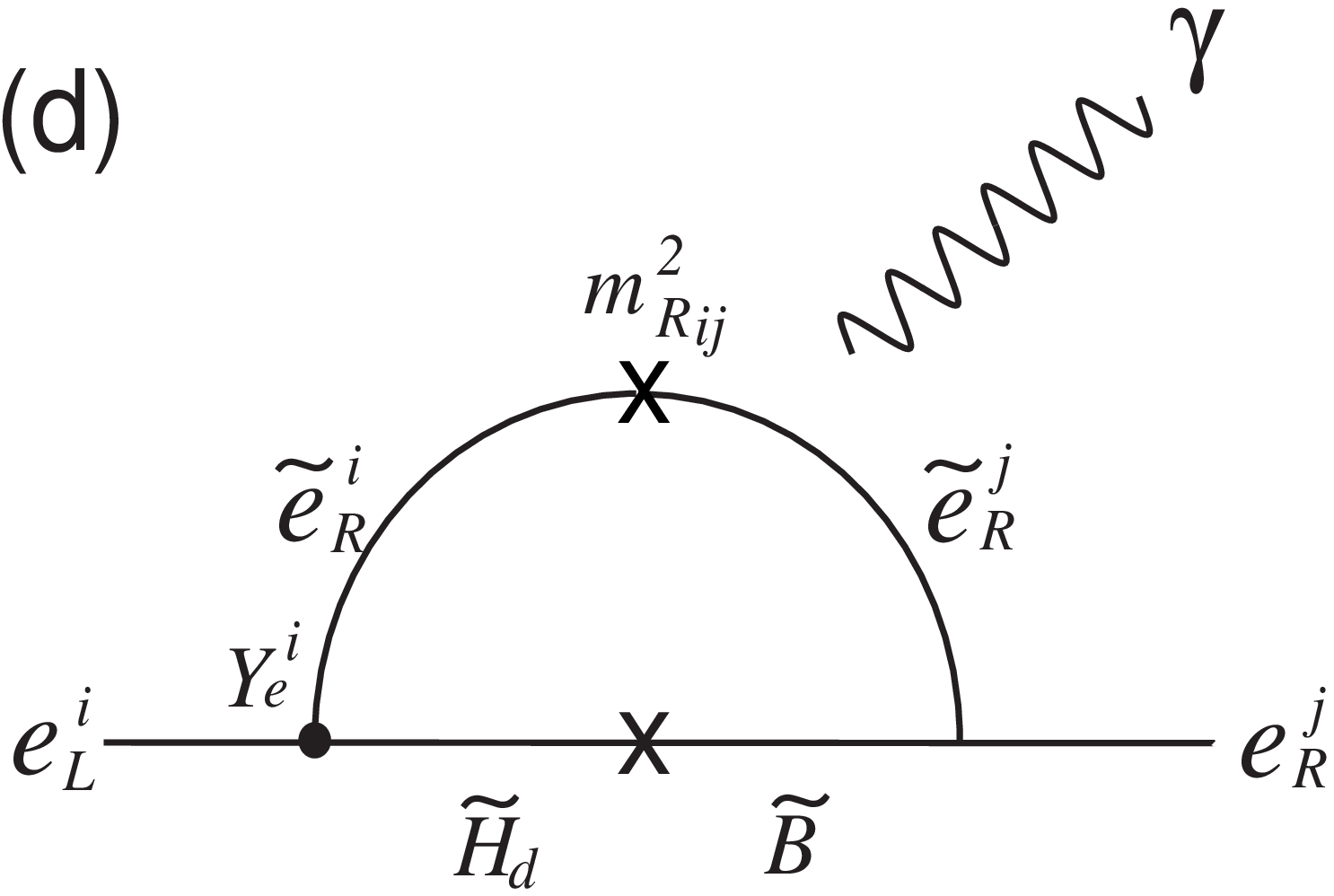}
 \caption{Chargino (a) and neutralino (b,c,d) loop diagrams to generate
         the  LFV processes. The off-diagonal elements of $m^2_L$ and $m^2_R$ come from
          $m^2_{\tilde \ell}$ and $m^2_{\tilde e}$, respectively.
      The mark $\bullet$ stands for a chirality flipping.
      The chirality flipping for Bino diagram (b) is given as
      $m^2_{LR} = A_e v_d - \mu\tan\beta \, M_e$.
          There are diagrams in which the chirality is flipped in the  external lines,
      but contributions from such diagrams are small.}
\label{diagram}
\end{figure}

In  SUSY models,  the LFV processes are described by loop diagrams.
The charginos and neutralinos propagate in the loop as shown in the
Fig.\ref{diagram}. When the SUSY breaking mass terms and the $A$-terms
violate lepton flavor, the branching ratio of the LFV processes can be
comparable to the experimental results. So the flavor structure of the
SUSY breaking parameters is constrained \cite{Gabbiani:1988rb}.

When the SUSY breaking masses are universal and the $A$-term
coefficient is proportional to the Yukawa coupling, there is no source
for any LFV. However, even if there exists no LFV source in  the SUSY
breaking parameters at the cutoff scale, the sources for LFV  can be
generated through the neutrino Dirac Yukawa couplings as long as the
coupling matrices are not proportional to the charged lepton Yukawa
matrix. The Majorana couplings can also generate LFV.

The RGEs above the scale of right-handed neutrino Majorana masses and
and triplet Higgs fields are written in a proper notation as
\begin{eqnarray}
(4\pi)^2 \frac{d}{d\ln Q} m_{\tilde \ell}^2
&\!\!=&\!\! \{Y_e Y_e^\dagger, m_{\tilde \ell}^2 \} +
  \{Y_\nu Y_\nu^\dagger, m_{\tilde \ell}^2 \}  + 3  \{f_L f_L^\dagger,m_{\tilde \ell}^2\}\\
&\!\!+&\!\! 2(Y_e m_{\tilde e}^2 Y_e^\dagger + m_{H_d}^2 Y_e Y_e^\dagger+ A_e A_e^\dagger)
+ 2(Y_\nu m_{\tilde \nu}^2 Y_\nu^\dagger + m_{H_u}^2 Y_\nu Y_\nu^\dagger+ A_\nu A_\nu^\dagger)
\nonumber \\
&\!\!+&\!\!
6 (f_L (m_{\tilde \ell}^2)^{\rm T} f_L^\dagger + f_L f_L^\dagger m_{\Delta_L}^2 + A_{f_L} A_{f_L}^\dagger)
\nonumber \\
&\!\!-&\!\! 8 (\frac14 g^{\prime 2} M_1^2 + \frac34 g_2^2 M^2_2) - g^{\prime 2} S \,, \nonumber \\
%
(4\pi)^2 \frac{d}{d\ln Q} m_{\tilde e}^2
&\!\!=&\!\! 2 \{Y_e^\dagger Y_e, m_{\tilde e}^2 \} +
   \{f_R^{ee\dagger} f_R^{ee},m_{\tilde e}^2\}+
   2 \{f_R^{\nu e \dagger} f_R^{\nu e},m_{\tilde e}^2\}\\
&\!\!+&\!\! 4(Y_e^\dagger m_{\tilde \ell}^2 Y_e + m_{H_d}^2 Y_e^\dagger Y_e + A_e^\dagger A_e)
\nonumber \\
&\!\!+&\!\!
2 (f_R^{ee\dagger} (m_{\tilde e}^2)^{\rm T} f_R^{ee} +
f_R^{ee\dagger} f_R^{ee} m_{\Delta_R^{--}}^2 + A_{f_R}^{ee\dagger} A_{f_R}^{ee})\nonumber \\
&\!\!+ &\!\!
4 (f_R^{\nu e\dagger} (m_{\tilde \nu}^2)^{\rm T} f_R^{\nu e} +
f_R^{\nu e\dagger} f_R^{\nu e} m_{\Delta_R^{-}}^2 + A_{f_R}^{\nu e\dagger} A_{f_R}^{\nu e})
\nonumber \\
&\!\!-&\!\! 8 g^{\prime 2} M_1^2 +2 g^{\prime 2} S \,, \nonumber
\end{eqnarray}
where $\{X,Y\} = XY+YX$ and $S$ is a trace of SUSY breaking masses with hypercharge weight.

As it has been emphasized, the SUSY breaking scalar masses are
universal due to the  geometrical setup. On the other hand, if the
$F$-terms of $U$-moduli are zero, the $A$-terms are proportional to
the Yukawa couplings. However, the non-zero values of $F^U$ provide a
source of LFV in the form as shown in the Eq.(\ref{U-moduli}). Let us
see the $U$-moduli contribution in the basis where the charged lepton
Yukawa matrix is diagonal: $U_L^e Y_e U_R^{e\rm T} = Y_e^{\rm diag}$.
Since $U_L^0 x^L = (0,0,\sqrt{a_1^2+b_1^2+c_1^2})^{\rm T}$, the
derivative of the rank 1 part $(Y_0)_{ij} = x^L_i x^R_j$ is written as
\begin{equation}
U_L^e (\partial_U Y_0) U_R^{e\dagger} =
V_L^e  \left(\begin{array}{ccc}
          0&0&{\tt x}\\ 0&0&{\tt x}\\ {\tt x}&{\tt x}& {\tt x}
          \end{array}
    \right) V_R^{e \dagger} \,,
\label{U-moduli-2}
\end{equation}
where the non-zero values in the entries are shown by $\tt x$. Note
that the (1,3) and (3,1) elements in the above matrix become zero when
$\tan \theta_s^{L,R}= 1$. Since the mixing angles in $V^e_{L,R}$ are
small, the elements $A_{ij}$ $(i,j \leq 2)$ can be small while the
elements $A_{i3}$ and $A_{3i}$ can be large.
In fact, the experimental bounds ${\rm Br} (\mu \rightarrow e\gamma) <
1.2 \times 10^{-11}$ \cite{Brooks:1999pu} and the EDM of the electron
$|d_e| < 1.6 \times 10^{-27}$ \cite{Regan:2002ta} provide the most
severe constraint on $A_{12}$ and Im$A_{11}$. Due to the structure of
Eq.(\ref{U-moduli-2}), $A_{i3}$ and $A_{3i}$ from the $U$ moduli
contribution can be the sources of LFV while keeping  the elements
$A_{12}$ and $A_{11}$  to be small. The $U$-moduli contribution can
generate the off-diagonal elements of SUSY breaking scalar mass
squared matrices through the RGEs.

In the minimal SUGRA, the non-proportionality of the $A$-term never
develops. In the intersecting D-brane models, the non-minimality of
the $A$-terms can be included when $F^U \neq 0$ while the SUSY
breaking scalar masses have degeneracy for different generations at the
cutoff scale due to the geometrical setup of D-branes.
Due to the particular form of the $U$-moduli contribution as shown in
the Eq.(\ref{U-moduli-2}), the (1,3) element can be larger than the
usual hierarchical assumptions for the non-minimal $A$-terms. The RGE
effects are not decoupled till the electroweak scale and due to this
the off-diagonal elements for both left- and right-handed slepton mass
matrices are generated. The
 right-handed off-diagonal elements  are always larger than the left-handed
 elements due
to a difference in the  coefficients of the terms involving $A_e$s in
the RGEs.

We enumerate the sources of LFV in the SUSY breaking scalar mass
matrices at the weak scale in the mSUGRA model as follows:
\begin{enumerate}
    \item Neutrino Dirac Yukawa coupling \cite{Borzumati:1986qx}

        The neutrino Dirac Yukawa coupling $Y_\nu$ can generate the off-diagonal
        elements of left-handed SUSY breaking slepton mass matrix
        $m^2_{\tilde \ell}\,$.
        Hence, the chargino contribution can dominate in the LFV processes.
        The RGE effects are decoupled at the right-handed neutrino Majorana mass scale.

    \item Majorana coupling for left- and right-handed leptons

        The left-handed Majorana coupling $f_L$ is needed in type II seesaw.
        The right-handed Majorana coupling $f_R$ also participates in the light
        neutrino mass  when $B-L$ charge is gauged.
        The Majorana coupling $f_L$ and $f_R$ can generate the off-diagonal
        elements of both left- and right-handed slepton mass matrices,
        $m^2_{\tilde \ell}$, $m^2_{\tilde e}$.
        The RGE effects are decoupled at the $\Delta_{L,R}$ mass scale,
        and the right-handed neutrino Majorana mass scale for $f_R^{\nu e}$ coupling.

    \item SU(5) GUT \cite{Barbieri:1994pv} or the left-right unification \cite{Babu:1999ge}

        Since the right-handed selectron can be unified in the $\bf 10$ dimensional representation
        of the $SU(5)$ grand unification,
        the off-diagonal elements of right-handed selectron can be generated above
        the unified scale.
        The generated off-diagonal elements of $m_{\tilde e}^2$ are related to the CKM mixings.
        In the left-right unified models, the two Higgs bidoublets are needed to
        generate the CKM mixings, and the two different Yukawa matrices are
         sources
        of off-diagonal elements for  both left- and right-handed sleptons.
        We do not discuss these sources in this paper.


\end{enumerate}

\section{Numerical studies}

In this section, we will show the numerical calculations of the
branching ratio of the LFV decays and the EDM of the electron in the
context of intersecting D-brane models.

We set up the parameters to show the numerical results as follows. The
charged lepton mass matrix is given as rank 1 matrix plus small
correction. The rank 1 matrix is given as Eq.(\ref{Yukawa0}) in the
basis where light neutrino mass matrix is diagonal. In the minimal
brane configuration, such as shown in the Fig.1, the parameters are
given \cite{Dutta:2005bb}
\begin{equation}
a_1 : b_1 : c_1 = \vartheta \left[ \begin{array}{cc} \varepsilon\\ 0 \end{array} \right](t):
 \vartheta \left[ \begin{array}{cc} -\frac13+\varepsilon\\ 0 \end{array} \right](t):
 \vartheta \left[ \begin{array}{cc} \frac13 + \varepsilon\\ 0 \end{array} \right](t)\,.
\end{equation}
For the calculation, we use $\varepsilon = 0.1$ and $t = 1.5$. Then
$\theta^L_a = 47^{\rm o}$ and $\theta^L_s = 37^{\rm o}$. For
simplicity, the Yukawa matrix is assumed to be symmetric. Then
$U$-moduli contribution of the $A$-term which is proportional to the
derivative of Yukawa coupling is calculated in the basis where the
charged lepton Yukawa matrix is diagonal
\begin{equation}
A_e^U = c \,A_0 \ V_L^e \left(\begin{array}{ccc}
          0&0&0.22\\ 0&0&0.26\\ 0.22&0.26& 0.84
          \end{array}
    \right) V_R^{e \dagger} \,,
\label{U-moduli-A}
\end{equation}
where $A_0$ is a dimensionful coupling coefficient and $c$ is a
coefficient. If $F^U=0$, $c=0$, the trilinear coupling is given as
$A_e = A_0 Y_e + A_e^U$. More precisely, the $U$-moduli derivative of
the correction matrix $\delta Y$ may  also contribute, but we neglect
its contribution here since its $U$-moduli derivative does not appear
to be large and has a model dependence. We will choose the mixing
angles in $V_L^e = V_R^e$ as $\theta_{12}^e = 0.1$, $\theta_{23}^e =
0.05$ and $\theta_{13}^e = 0.005$. There can be 5 phases in the
unitary matrix $V_L^e$ up to an overall phase in general, but for
simplicity, we assume that there is no CP phase in $V_L^e$. The
neutrino mixing angles are given in the Eq.(\ref{neutrino-mixing}). In
the choice above, $U_{e3} = 0.07$ and $\theta_{\rm sol} = 33^{\rm o}$.

The neutrino Dirac coupling can be written as $Y_\nu = V_L^{e\nu}
Y_\nu^{\rm diag} V_R^{e\nu\dagger}$   in the basis where the
charged-lepton Yukawa coupling is diagonal. In a general scheme, the
unitary matrix $V_L^{e\nu}$ is completely free. For example,
$V_L^{e\nu *}$ is the MNSP matrix when type I seesaw is dominated and
the right-handed neutrino Majorana mass matrix is proportional to
identity matrix. However, in the present scheme of ``almost rank 1
Yukawa matrices", the unitary matrix $V_L^{e\nu}$ is close to an
identity matrix like the CKM matrix, and $V_L^{e\nu} \simeq V_L^e$
when the Dirac Yukawa coupling $Y_\nu$ is hierarchical like up-type
quark masses. We will use $V^{e\nu}_{L,R}=V^e_L$ to express the
numerical result. As we have already noted, we use $\theta_{12}^e =
0.1$, $\theta_{23}^e = 0.05$ and $\theta_{13}^e = 0.005$.

\subsection{LFV decays}

\begin{figure}[tb]
 \center
 \includegraphics[width=10cm]{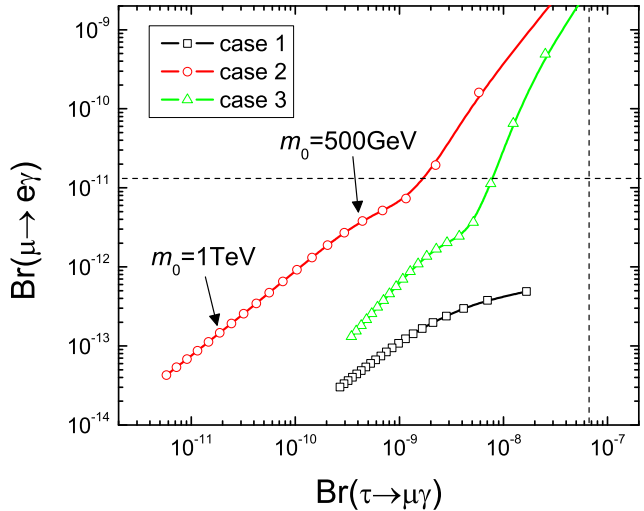}
 \caption{The branching ratios of the LFV decays are plotted.
 In the plot, slepton mass at cutoff scale is varied with 50 GeV steps.
 The detailed parameters we used are given in the text.
 Dashed lines are drawn for the current experimental bounds at 90\%CL.
 The lines are plotted for the following cases:
 (Case 1) $Y_{\nu_\tau} = Y_t$ and $c=0$,
 (Case 2) $Y_{\nu_\tau} = 0.1\, Y_t$ and $c=0.1$,
 (Case 3) $Y_{\nu_\tau} = Y_t$ and $c=0.1$.
 $c$ is the coefficient given in the Eq.(\ref{U-moduli-A}).}
\label{mue-taumu}
\end{figure}

We plot the branching ratios of $\tau \rightarrow \mu\gamma$ and $\mu
\rightarrow e\gamma$ in Fig. \ref{mue-taumu}. For the SUSY breaking
parameters, we assume that $m_{\tilde \ell}^2{}_{ij} = m_{\tilde
e}^2{}_{ij} = m_0^2 \,{\bf 1}$ and $A_e = A_0 Y_e + A_e^U$ at the
cutoff scale $M_*$ as we have mentioned. In the intersecting D-brane
models, the SUSY breaking scalar mass is not necessarily universal for
different representation, though the flavor degeneracy is achieved. We
assume that the left and right scalar masses to be same just for
simplicity. The cutoff scale is related to the string scale and the
volume of the extra dimensions. We choose that $M_* = 10^{17}$ GeV in
the calculation. We take gaugino mass $M_{1/2} = 500$ GeV at $M_*$,
$A_0 = 500$ GeV and Higgsino mass $\mu = 500$ GeV (we choose the
signature of $\mu$ to make the SUSY contribution of anomalous magnetic
moment of muon \cite{Brown:2001mg} to be positive). The value of
$\tan\beta$ which is the ratio of the vacuum expectation values for
Higgs fields is taken to be $\tan\beta =50$. The amplitudes for the
LFV decays are naively proportional to $\tan\beta$, thus the branching
ratios are $\propto \tan^2\beta$. In the Fig.\ref{mue-taumu}, we vary
 $m_0$ in 50 GeV steps and the maximal value is $m_0 = 1250$ GeV
(which corresponds to the lightest stau of about 1 TeV). In order to
show the results clearly, we assume that the LFV sources are only the
Dirac neutrino Yukawa coupling and the $U$-moduli contribution in
$A_e$. We neglect the sources arising in the Majorana couplings
(Eq.(\ref{Majorana})) by assuming them to be small. In the plot, Dirac
neutrino coupling is the only source in the case 1. We take the
largest right-handed Majorana mass to be $10^{15}$ GeV. In the case 2,
the $U$-moduli contribution is the dominant source of LFV. The case 3
has both sources. It is easy to see that the  neutrino Dirac couplings
makes the branching ratio Br($\tau\rightarrow \mu\gamma$) large. This
is because that these couplings
 generate the off-diagonal (2,3) element of
the left-handed slepton mass matrix and contributes to the chargino
diagram of $\tau\rightarrow \mu\gamma$. The flavor violation source
arising from the  neutrino Dirac couplings can contribute to the
$\mu\rightarrow e\gamma$ decay since the (1,3) element is also
generated, but this element is smaller than the (2,3) element. If we
switch on the $U$-moduli contribution, the (1,3) element can be
comparable to the (2,3) element and
thus, the $U$-moduli contribution increases the $\mu \rightarrow
e\gamma$ decay rate  more than the $\tau \rightarrow \mu\gamma$ decay
rate. The reason for the behavior of the lines being different (in the
Fig.\ref{mue-taumu}) when $m_0$ is smaller than 400 GeV is that the
Bino-Bino diagram dominates rather than the chargino diagram due to
the large left-right mixings of the slepton. The qualitative behaviors
of the cases 1 and 2 are not very different even if we change the
numerical parameters, but the case 3 depends much on the initial
condition such as $A_0$ and $\mu$ since there can be a slight
cancellation among the diagrams. The right- and left-handed lepton
decays do not have interference, and a huge cancellations among the
diagram for the branching ratios  happen hardly.

\begin{figure}[tb]
 \center
 \includegraphics[width=8cm]{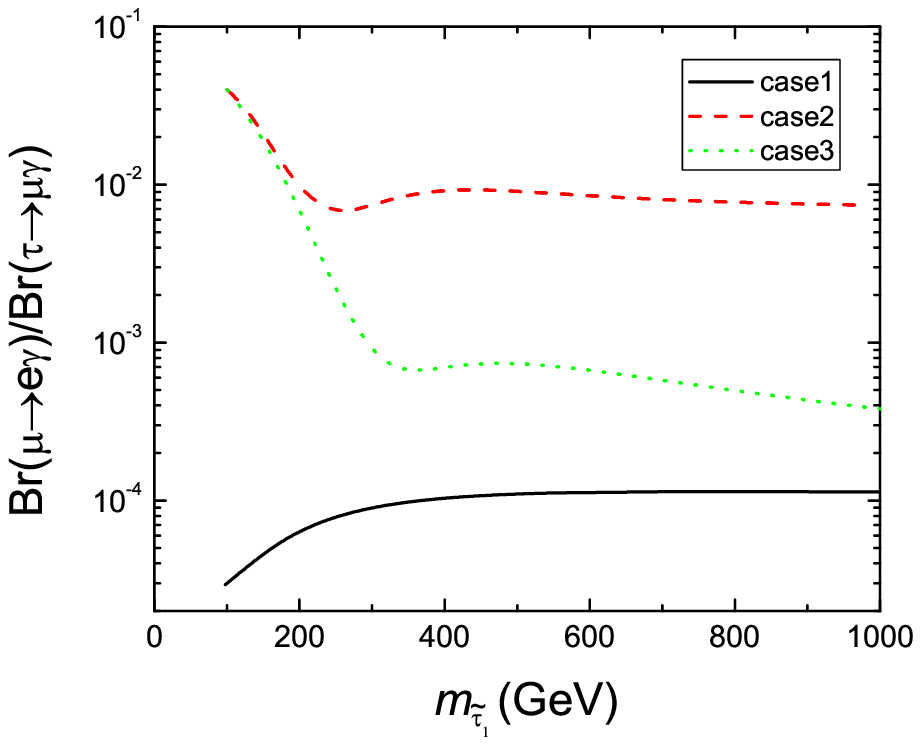}
 \includegraphics[width=8cm]{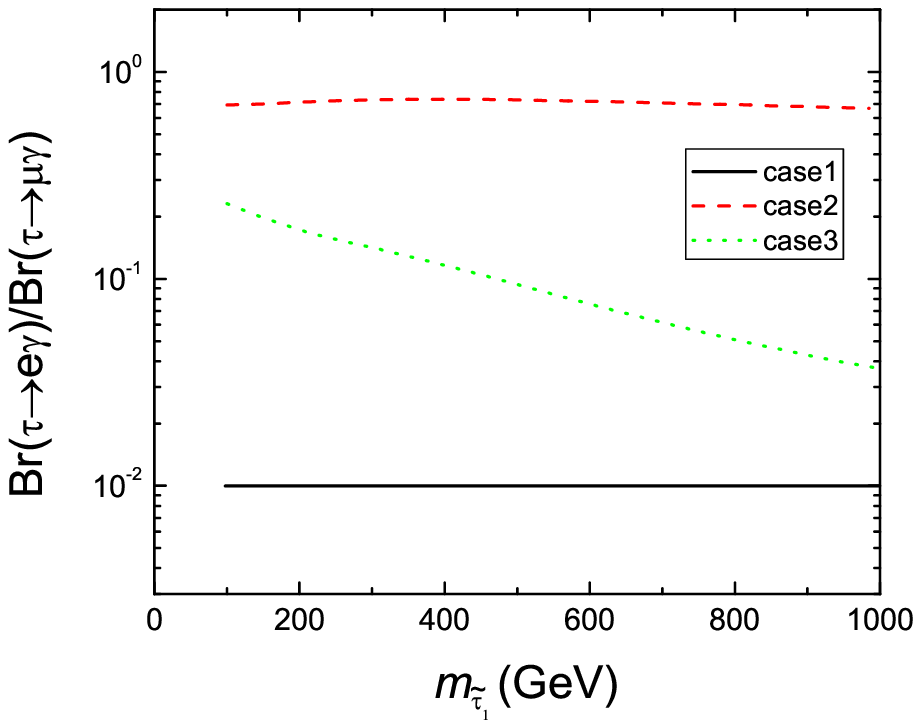}
 \caption{
 Plots for the ratio of branching ratios.
 The  lines are plotted for the following cases:
 (Case 1) $Y_{\nu_\tau} = Y_t$ and $c=0$,
 (Case 2) $Y_{\nu_\tau} = 0.1\, Y_t$ and $c=0.1$,
 (Case 3) $Y_{\nu_\tau} = Y_t$ and $c=0.1$.
 }
\label{ratio}
\end{figure}

The branching ratio for each decay mode depends on the initial
conditions. However, as shown in the Fig.\ref{ratio}, the ratio of the
branching ratios can be a good prediction for different LFV sources.
In the figure, we use the same initial conditions as before. The ratio
of the branching ratio is almost determined by the mixing angles in
$V_L^{e\nu}$ for the case 1 and the ratio of the (1,3) and the (2,3)
elements in the $A_e^U$ for the case 2. Therefore, if all the
branching ratios
 for $\tau
\rightarrow \mu\gamma$, $\tau \rightarrow e\gamma$ and $\mu
\rightarrow e\gamma$ are measured, we can obtain important
information to identify the LFV source.
In fact, the following relations are satisfied approximately:
Br($\mu\rightarrow e\gamma$)/Br($\tau\rightarrow \mu\gamma)
\sim (\theta_{13}^{e})^2$/Br($\tau\rightarrow \mu\bar\nu_\mu\nu_\tau)$
 for the case 1,
$\sim (A_{e}{}_{13}/A_e{}_{33})^2$/Br($\tau\rightarrow \mu\bar\nu_\mu\nu_\tau$)
 for the case 2 and
 Br($\tau\rightarrow e\gamma$)/Br($\tau\rightarrow \mu\gamma)
\sim (\theta_{13}^{e}/\theta_{23}^e)^2$ for the case 1, $\sim
(A_e{}_{13}/A_e{}_{23})^2$ for case 2. In the case 3, the LFV sources
are mixed, we do not have such simple expressions. These ratios of the
branching ratios do not depend much on the initial conditions such as
$m_0$, $A_0$, $M_{1/2}$, $\mu$ and $\tan\beta$ if the chargino diagram
provides the  dominant contribution. When the sleptons are light and
the left-right mixing becomes large, the Bino diagram can contribute
to $\mu \rightarrow e\gamma$ and bends the lines for the ratio
Br($\mu\rightarrow e\gamma$)/Br($\tau\rightarrow \mu\gamma$) in the
Fig.\ref{ratio} for smaller $m_0$.

The large Majorana couplings $f$ can contribute to the LFV decays due
to its off-diagonal terms. If the type II seesaw dominates the
neutrino masses, the ratios of the branching ratios are almost
determined by the neutrino mixings when the Majorana couplings are the
dominant sources of the LFV decays.
In this case, the ratio
Br$(\mu\rightarrow e\gamma)$/Br$(\tau\rightarrow \mu\gamma$)
is about $U_{e3}^2$/Br($\tau\rightarrow \mu\bar\nu_\mu\nu_\tau$)
while the ratio
Br$(\tau\rightarrow e\gamma)$/Br$(\tau\rightarrow \mu\gamma)$
is about $(U_{e3}/U_{\mu 3})^2 \simeq (\theta_{12}^e)^2$.
The first value is similar to the pure $U$-moduli case (case 2) and
the second value is similar to the Dirac neutrino case (case 1). Thus
the observation of the ratios can sort out the LFV sources.

Usually, the 13 mixing is smaller than the 23 mixing in $V_L^{e\nu}$
or $U_L^e$, even if we use a different setup, and thus the
$\tau\rightarrow e\gamma$ decay rate is expected to be smaller than
the $\tau\rightarrow \mu\gamma$ decay rate. However, if the $U$-moduli
contribution  dominates, those two decay rates can be comparable since
the (1,3) and the (2,3) elements in $A_e^U$ are comparable. So
measuring the ratio Br$(\tau\rightarrow e\gamma)$/Br$(\tau\rightarrow
\mu\gamma)$ is very important to see the presence of the $U$-moduli
contribution.

\subsection{EDM}

The other important observables to select the sources of LFV are the
EDMs of the electron and the muon. If the trilinear scalar coupling
$A_0$ and the Higgsino mass $\mu$ are complex parameters, the EDMs can
be large even if we do not have any source of LFV violation. However,
if those are complex in general, the EDM of the electron can be too
large compared to the experimental bounds when the slepton masses are
less than around 1 TeV \cite{Franco:1983xm}. Thus a cancellation is
needed in that case to satisfy the bound \cite{Ibrahim:1998je}. It is
unnatural to have cancellations for both the electron and the muon and
thus the muon EDM will be large enough to be detected in the  future
experiments~\cite{Semertzidis:1999kv}. It is often assumed that the
$A_0$ and $\mu$ are real to satisfy the experimental bound naturally.
In this case, the amount of EDMs are related to the source of LFV.

\begin{figure}[tb]
 \center
 \includegraphics[width=10cm]{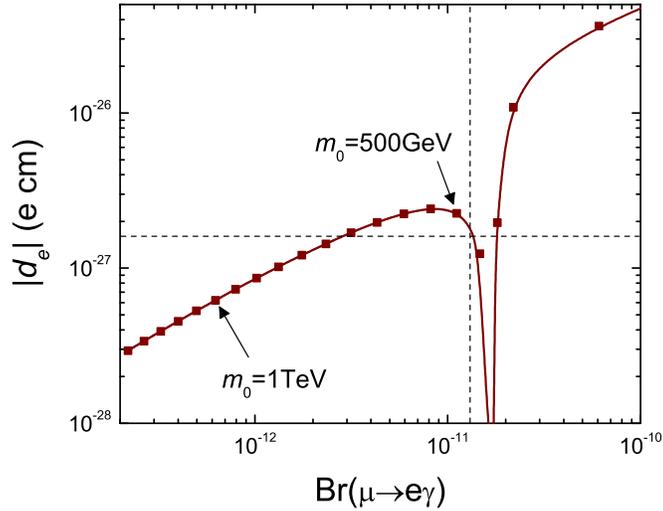}
 \caption{Plots for electron EDM and branching ratio of
 $\mu\rightarrow e\gamma$.
 In the plot, the slepton mass at the cutoff scale is varied with 50 GeV steps.
 Dashed lines are drawn for the current experimental bounds at 90\%CL.}
\label{mue-edm}
\end{figure}

Let us suppose that $A_0$ and $\mu$ are real and all the CP phases are
in the Yukawa couplings. The EDMs are imaginary part of the amplitude
of the loop diagram. Since in the diagram, where the chirality
flipping vertex does not include  CP phase in generation mixings, the
Bino-Bino diagram dominates the EDM calculation. The imaginary parts
of $(A_e)_{11}$ and $(A_{e})_{12} m_{\tilde e}^2{}_{21}$ etc in the
basis where the charged-lepton matrix is real and diagonal can be also
important. However, if the $A$-term is proportional to the Yukawa
coupling, such imaginary parts are small.
The electron EDM can be proportional to $\mu \tan \beta \, m_\tau$
when the (1,3) mixings for both left- and right-handed slepton mass
matrices are generated. If the neutrino Dirac Yukawa coupling is the
only source of  LFV, the off-diagonal elements of right-handed slepton
mass matrix are very small and consequently, the EDM of the electron
is small, $d_e \sim 10^{-33}\, e\,$cm. Even in the type I seesaw with
generic right-handed Majorana mass matrix, the electron EDM is at most
$d_e \sim 10^{-29}\, e\,$cm \cite{Ellis:2002fe}. If the $U$-moduli is
a source of LFV processes, the off-diagonal elements for both left-
and right-handed slepton mass matrices can be generated  and thus the
electron EDM can be enhanced to reach the current experimental bound
$|d_e| < 1.6 \times 10^{-27} \,e\,$cm \cite{Regan:2002ta}. Hence, the
electron EDM is an important observable to see whether the LFV arises
from only the neutrino Dirac Yukawa coupling or not.

In the intersecting D-brane models, if we assume that the $F$-term of
moduli does not have any phase then $A_0$ does not get any phase. The
phase of the Higgsino mass depends on the model, but it may be related
to the SUSY breaking  parameters and thus can be real in such an
assumption. On the other hand, the Yukawa couplings can be complex if
we include the Wilson line phases in the theta function
\cite{Cremades:2003qj}. Then the $U$-moduli contribution of the $A$
term can be generically complex while $A_0$ is real.

We plot the electron EDM and the branching ratio of $\mu\rightarrow
e\gamma$ in the Fig.\ref{mue-edm} in the case where the $U$-moduli
contribution is large enough using the same input for the SUSY
breaking parameters as before. In this case, the values different from
what has be shown for the neutrino Dirac Yukawa couplings do not
change the plot very much since the chargino diagram does not
contribute to the EDM. In general each component of $A_e^U$ can be
complex independently. In the plot, we
 take the overall factor for the $U$ moduli contribution $c$ to be pure
imaginary for simplicity. The EDM  can easily saturate the current
experimental bound. For larger $m_0$, the $A_e v_d$ contribution is
larger compared  to  the $\mu \tan\beta\, m_\tau$ part in the
left-right slepton mixing while for smaller $m_0$ ($<400$ GeV), the
$\mu\tan\beta\, m_\tau$ contribution becomes larger than the $A_e v_d$
part. This is because $\mu \tan\beta\,m_\tau$ contribution needs
triple mass insertion in the Bino diagram and thus the amplitude is
suppressed by larger power of $m_0$ than the $A_e v_d$ contribution
which can be produced by a single mass insertion.

The presence of large complex Majorana couplings can also saturate the
EDM bound when $\mu\tan\beta \,m_\tau$ contribution is large for light
sleptons. The $A_e v_d$ part is small when $U$-moduli contribution is
absent. When sleptons are heavy (depending on $\mu\tan\beta$), the
$\mu\tan\beta \,m_\tau$ contribution is suppressed due to the triple
mass insertion and thus the EDM becomes smaller for a fixed
Br$(\mu\rightarrow e\gamma)$, comparing to the case when the
$U$-moduli contribution is dominant.

The muon EDM is $d_\mu \sim 10^{-26} - 10^{-24}
\,e\,$cm as long as the bound for $d_e$ is satisfied and there is no
huge cancellation in $d_e$. The ratio of the EDMs
does not depend on the ratio of the corresponding charged lepton
masses, $d_\mu/d_e \neq m_\mu/m_e$.

\section{Conclusions}

We have discussed the flavor sector in the
intersecting D-brane models.
In the D-brane models,
the low energy effective action for the zero modes
such as particles in MSSM
can be calculated using the geometrical parameters.
The neutrino mixing is elegantly realized in the context of the almost
rank 1 Yukawa matrix. The flavor degeneracy of SUSY breaking scalar
masses are realized when the generation is simply replicated at the
intersection of the D-branes. The non-proportional part of the scalar
trilinear coupling is  obtained
 when the $F$-term of the $U$-moduli is not zero.

Emphasizing  the flavor degeneracy of SUSY breaking scalar masses in
these models, we study the lepton flavor violating processes. The
$U$-moduli contribution of the scalar trilinear coupling can be the
source of lepton flavor violation as well as the neutrino Dirac and
Majorana Yukawa couplings which are included in the MSSM plus
right-handed neutrino.

We calculate the branching ratios of  the LFV decays and the
EDM of the electron $d_e$. The observations of the $d_e$ and
the ratio of the branching ratios for $\mu \rightarrow e\gamma$, $\tau
\rightarrow \mu\gamma$ and $\tau\rightarrow e\gamma$ are important to
sort out the sources of LFV as shown in the
Fig.\ref{ratio}.

The  bound for the EDM of the electron can be improved to $d_e \sim
10^{-32} \,e\,$cm in the planned experiments \cite{Lamoreaux:2001hb}.
Under the assumption that the SUSY breaking parameters and the
Higgsino mass are real, it is possible to see whether we have any
source of LFV in addition to the Dirac neutrino Yukawa couplings in
mSUGRA. The $U$-moduli contribution in our scheme can saturate the
current experimental bounds for $d_e < 1.6 \times 10^{-27}\,e\,$cm as
well as the Br($\mu \rightarrow e\gamma$). The current bound for the
branching ratio is Br($\mu\rightarrow e\gamma) < 1.2 \times 10^{-11}$
and it can go down to $\sim 5 \times 10^{-14}$ in the near future
\cite{Grassi:2005ac}.

The $U$-moduli contribution in the trilinear scalar coupling can make
Br($\tau \rightarrow e\gamma$)/Br($\tau \rightarrow \mu\gamma$) to be
order 1. In the models where the Dirac neutrino or the Majorana
couplings are the primary sources of LFV, this ratio is much smaller
than 1.
  In order to completely sort
out the LFV sources from the ratio of the branching ratios,  the
$\tau$ LFV decays with branching ratio $10^{-9}-10^{-10}$ need to be
at least measured. At present, the upper bound on the branching ratio
of  $\tau \rightarrow \mu\gamma$ is $6.8 \times 10^{-8}$
\cite{Brooks:1999pu}
 and this bound can be improved
to $10^{-10}$ at the ILC-Super B \cite{new}.
%

%

\section*{Acknowledgments}
We would like to thank Justin Albert for providing us the information
on the future limit of the $\tau\rightarrow\mu\gamma$ branching ratio
at the ILC-Super B.

\end{document}